\documentclass[letterpaper, 10 pt, conference]{ieeeconf}

\IEEEoverridecommandlockouts
\usepackage{amsmath,amssymb,amsfonts}
\usepackage{mathtools}
\usepackage{graphicx}
\usepackage{textcomp}
\usepackage{xcolor}
\usepackage{cite}
\usepackage{algorithm}
\usepackage{algorithmic}
\usepackage{bm}
\usepackage[hidelinks]{hyperref}
\usepackage{subcaption}
\usepackage{tikz}
\usetikzlibrary{arrows.meta,positioning,calc,shapes}

\DeclareMathOperator{\diag}{diag}
\DeclareMathOperator{\cond}{cond}
\DeclareMathOperator{\tr}{tr}
\newcommand{\R}{\mathbb{R}}
\newcommand{\Sbb}{\mathbb{S}}

\newtheorem{theorem}{Theorem}
\newtheorem{proposition}{Proposition}
\newtheorem{lemma}{Lemma}

\newtheorem{definition}{Definition}
\newtheorem{remark}{Remark}

\begin{document}

\title{Learning Integral Quadratic Constraints\\with Applications to Observer Design}

\author{Mario Virdis, Florian Dörfler, Paal E. Engelstad, Mathias Hudoba de Badyn%
\thanks{M. Virdis, P. E. Engelstad, and M. Hudoba de Badyn are with the Department of Technology Systems, University of Oslo, Kjeller, Norway. E-mails: \texttt{\{mario.virdis, paal.engelstad, mathihud\}@its.uio.no}}%
\thanks{F. Dörfler is with the Automatic Control Laboratory, ETH Zürich, Switzerland. E-mail: \texttt{dorfler@ethz.ch}}%
\thanks{This work was supported by the Research Council of Norway under the Centre for Space Systems and Sensors (309835).}%
}%

\maketitle

\begin{abstract}
Existing methods for constructing nonlinear neural operators that satisfy incremental integral quadratic constraints (IQCs) by construction require the IQC multiplier to be fixed before training.
This forces an a priori commitment to a specific trade-off between contraction rate and expressivity of the model.
We propose to jointly learn the IQC multiplier, the neural operator, and, when the system matrix depends affinely on a design parameter such as an observer gain, the design parameter itself, all while maintaining contraction guarantees.
The main result is a parametrization of the IQC multiplier, which renders it a differentiable function of unconstrained parameters and eliminates the need for online semidefinite program solves.
We validate the approach on an observer design problem, which demonstrates that the proposed method improves approximation quality over fixed-multiplier designs while preserving certified contraction of the observer.
\end{abstract}

\section{Introduction}
\label{sec:intro}

The integral quadratic constraint (IQC) framework~\cite{megretski_system_1997,scherer_dissipativity_2022,seiler_stability_2015,veenman_robust_2016} provides a powerful tool for certifying robust stability of interconnections between dynamical systems and nonlinear operators.
Its extension to the incremental setting~\cite{wang_robust_2019,forni_differentially_2013} enables contraction analysis via the feasibility of a dissipation linear matrix inequality (LMI).
For systems with polynomial dynamics, this condition takes the form of a pointwise matrix inequality that can be verified via sum-of-squares (SOS) programming~\cite{lohmiller_contraction_1998,manchester_control_2017}.
A natural synthesis question is how to construct neural operators that satisfy the corresponding dissipation inequality \emph{by construction}, while retaining sufficient expressivity to approximate a desired target behavior.

Recent works have addressed this through direct parametrizations of incrementally dissipative operators, most notably the recurrent equilibrium network (REN) framework~\cite{revay_lipschitz_2020,revay_recurrent_2024}.
RENs provide an unconstrained parametrization for a class of discrete-time operators that satisfy an incremental IQC with a \emph{fixed} multiplier $\Pi$.
For example, the Youla-REN architecture~\cite{wang_youla-ren_2022} and neural system level synthesis~\cite{furieri_neural_2022} have leveraged this property to design neural feedback control policies with certified contraction and stability guarantees.

In a standard REN application, the IQC multiplier, which encodes what behavior the operator is \emph{allowed} to exhibit, must be fixed before training begins.
The multiplier determines both the contraction rate $\beta$ (convergence speed) and the admissible gain $\rho$, the incremental gain the neural operator may have in every direction, which are locked into a trade-off described by a Pareto front in $(\beta,\rho)$-space.
Fixing the multiplier confines the operator to a single point in this space, chosen without knowledge of the task loss landscape.
If the chosen multiplier is too conservative, the neural operator lacks the expressivity to fit the target; if too permissive, it sacrifices contraction margins unnecessarily.

In this paper, we formulate and solve an \emph{IQC learning problem}, in which the multiplier is not prescribed but learned by the training process jointly with the rest of the operator's parameters.
The central question becomes \emph{which IQC the operator should satisfy}, and the answer is determined by the task objective, not by the designer ad hoc.

Alternative approaches require solving a semidefinite program (SDP) at every gradient step, e.g.~\cite{pauli_training_2022}, which is expensive for large models, or after training \cite{fazlyab_safety_2022}, which does not allow robust synthesis.
We address this problem by exploiting the property that the set of solutions of LMI certificates is convex, so a finite collection of certificates computed before training can be smoothly interpolated to produce a continuously parametrized family of feasible multipliers.
By mapping unconstrained parameters $\zeta\in\R^N$ onto the probability simplex, the IQC multiplier becomes a differentiable function of $\zeta$, enabling joint optimization of multiplier and operator via standard gradient descent.
Furthermore, when the closed-loop system matrix depends affinely on a design parameter, such as an observer gain $L$, a change-of-variable reformulation renders the LMI affine in all decision variables, enabling the co-optimization of both the multiplier and the design parameter through the same interpolation mechanism.
In the context of this paper, we understand \emph{learning} as the unconstrained minimization of a non-convex cost function (typically mean square error (MSE)), over a dataset, via gradient descent, where the decision variables are the parameters of the neural operator.
Optimality is thus only attained locally, as is usual for optimization problems involving neural operators.

\subsection{Contributions}

The main contributions of this paper are as follows.

\begin{enumerate}
	\item We formulate the \emph{IQC learning problem} for nonlinear incrementally-constrained operator design, that is, jointly learning the IQC multiplier and the operator parameters so as to minimize a task objective while maintaining certified dissipation. This differs from existing direct parametrizations~\cite{revay_recurrent_2024,wang_youla-ren_2022}, where the multiplier is fixed a priori (Section~\ref{sec:certificate_interp}).

	\item We prove that convex interpolation over a finite set of certificates, combined with a smooth simplex mapping, yields a differentiable parametrization of the IQC multiplier that maintains LMI feasibility for all parameter values. When the system matrix depends affinely on a design parameter, a reformulation into an LMI synthesis problem enables its co-optimization (Proposition~\ref{prop:main}, Section~\ref{sec:main_result}).

	\item We specialize the method to nonlinear observer design through the \emph{ICON} (Incrementally Constrained Observer with Neural dynamics) pipeline, where the learned certificate includes the observer gain alongside the IQC multiplier, producing a jointly trainable system with certified contraction (Section~\ref{sec:icon_method}).

	\item We demonstrate on a discrete-time observer problem that learning the IQC and observer gain jointly with the neural operator improves the approximation quality over fixed-multiplier designs, and that the optimizer autonomously selects a $(\beta,\rho)$ trade-off from the task loss alone (Section~\ref{sec:examples}).
\end{enumerate}

This paper is organized as follows: \S\ref{sec:prelim} introduces the observer error dynamics, dissipation inequalities, and the $(\beta,\rho)$ Pareto front. \S\ref{sec:certificate_interp} develops the differentiable IQC multiplier parametrization via convex interpolation and smooth simplex mapping. \S\ref{sec:main_result} states the main result guaranteeing LMI feasibility for all parameter values. \S\ref{sec:icon_method} assembles these components into the ICON pipeline for jointly training the neural operator, the IQC multiplier, and the observer gain. \S\ref{sec:examples} validates the approach on an anisotropic observer design problem, and \S\ref{sec:conclusion} concludes the paper.

\subsection{Notation}

We denote by $\Sbb^n$ the set of real symmetric $n\times n$ matrices, and with $\Sbb^n_+$ the set of positive definite symmetric matrices.
For $A,B\in\Sbb^n$, $A\succ B$ (resp.\ $A\succcurlyeq B$) denotes positive (semi-)definiteness of $A-B$.

\section{Preliminaries and Motivating Application}
\label{sec:prelim}

\subsection{Observer Error Dynamics}
\label{sec:setup}

Consider a discrete-time plant
\begin{equation}
	x_{k+1} = f(x_k) + g(x_k), \quad y_k = h(x_k),
	\label{eq:plant}
\end{equation}
where $x_k\in\R^n$ is the state, $y_k\in\R^{n_y}$ is the measured output, $f:\R^n\to\R^n$ is a known (possibly nonlinear) nominal dynamical map, $g:\R^n\to\R^n$ is a nonlinear unknown term, and $h:\R^n\to\R^{n_y}$ is the measurement function.
A Luenberger-type observer augmented with a neural operator $\phi_\theta$, where $\theta\in\mathbb{R}^{n_\theta}$ represents its parameters (sometimes omitted for readability), trained to approximate the unknown $g$ from data, takes the form
\begin{equation}
	\hat{x}_{k+1} = f(\hat{x}_k) + \phi(\hat{x}_k) + L\bigl(y_k - h(\hat{x}_k)\bigr),
	\label{eq:observer}
\end{equation}
where $L\in\R^{n\times n_y}$ is the observer gain.
Since $g$ is a memoryless function of the state, which $\phi_\theta$ approximates from data, we restrict $\phi$ to memoryless neural operators throughout the paper.

We define the estimation error as $e_k \coloneqq x_k - \hat{x}_k$.
Subtracting~\eqref{eq:observer} from~\eqref{eq:plant}, and substituting the relation
\begin{equation}
	g(x_k)-\phi(\hat{x}_k)=g(x_k)-\phi(x_k)+\phi(x_k)-\phi(\hat{x}_k),
\end{equation}
yields
\begin{equation}
\begin{split}
	e_{k+1} =& f(x_k)-f(\hat{x}_k)-L\,(h(x_k)-h(\hat{x}_k))\\
	&+\phi(x_k)-\phi(\hat{x}_k)+\underbrace{g(x_k)-\phi(x_k)}_{d_k}.
\end{split}
\end{equation}
The estimation error at time $k+1$ is thus the difference of the right-hand side of~\eqref{eq:observer} evaluated at $x_k$ and at $\hat{x}_k$, plus the additional term $d_k$, which depends only on the true trajectory.
The property of $d_k$ that the argument uses is its independence from $\hat{x}_k$, so that it influences the error recursion without feeding back into it.
Convergence of $e_k$ is therefore determined by the differential dynamics of the observer,
\begin{equation}
	\delta\hat{x}_{k+1} = F(\hat{x}_k)\,\delta\hat{x}_k + \delta v_k, \qquad \delta v_k = \frac{\partial \phi}{\partial x}(\hat{x}_k)\,\delta\hat{x}_k,
	\label{eq:variational}
\end{equation}
where $F(\hat{x}_k) \coloneqq \frac{\partial f}{\partial x}(\hat{x}_k) - L\frac{\partial h}{\partial x}(\hat{x}_k)$ is the estimate-dependent closed-loop Jacobian and $\delta v_k$ is the differential output of $\phi_\theta$.
Contraction of~\eqref{eq:variational} in turn implies that $\lVert e_k \rVert \to 0$ exponentially, up to an ultimate bound determined by $\sup_k \lVert d_k \rVert$~\cite{lohmiller_contraction_1998}.

\subsection{Dissipation Inequalities and IQCs}
\label{sec:iqc}

We adopt the dissipation-based formulation of IQC analysis, following~\cite{scherer_dissipativity_2022,seiler_stability_2015}, in its incremental form~\cite{forni_differentially_2013,wang_robust_2019,verhoek_convex_2023}.
Throughout, given any two admissible trajectories for a system, we denote the incremental signals $\delta \hat{x}_k = \hat{x}_k - \hat{x}'_k$, $\delta z_k = z_k - z'_k$, and $\delta v_k = v_k - v'_k$, where $z_k$ is the input and $v_k$ the output of the neural operator $\phi$.

\begin{definition}
	\label{def:iqc}
	An operator $\phi$ satisfies a static incremental IQC if, for all pairs of input-output trajectories,
	\begin{equation}
		\sum_{k=0}^T \begin{bmatrix} \delta v_k \\ \delta z_k \end{bmatrix}^\top
		\underbrace{\begin{bmatrix} Q_\phi & S_\phi \\ S_\phi^\top & R_\phi \end{bmatrix}}_{\Pi_\phi}
		\begin{bmatrix} \delta v_k \\ \delta z_k \end{bmatrix} \geq 0, \quad \forall\, T\geq 0.
		\label{eq:iqc_condition}
	\end{equation}
\end{definition}
We call $\Pi_\phi$ the (quadratic incremental static IQC) \emph{multiplier}.
When $\phi_\theta$ is implemented as a REN~\cite{revay_recurrent_2024}, it is possible to enforce a prescribed multiplier $\Pi_\phi$, such that for every $\theta\in\mathbb{R}^{n_\theta}$, $\phi_\theta$ satisfies an incremental IQC with the prescribed \emph{fixed} multiplier~$\Pi_\phi$.

We denote by $\bm{\Phi}(\Pi_\phi)$ the class of all operators satisfying the incremental IQC defined by~$\Pi_\phi$.
The class $\bm{\Phi}(\Pi_\phi)$ is substantially larger than the family of RENs parametrized by $\theta$.
Contraction of the augmented observer~\eqref{eq:observer} can then be certified via a dissipation LMI that incorporates the multiplier.
\begin{proposition}
	\label{prop:full_contraction}
	Consider the Luenberger-type observer~\eqref{eq:observer} with $\phi \in \bm{\Phi}(\Pi_\phi)$.
	The observer is contracting at rate $\beta\in(0,1]$ if there exists an incremental \emph{storage function} $V(\delta \hat{x}_k) = \delta \hat{x}_k^\top P\,\delta \hat{x}_k$, with $P\succ 0$, such that
	\begin{equation}
		V(\delta \hat{x}_{k+1}) \leq \beta\, V(\delta \hat{x}_k) + \begin{bmatrix} \delta z_k\\ \delta v_k \end{bmatrix}^\top
		\underbrace{\begin{bmatrix} Q & S \\ S^\top & R \end{bmatrix}}_{\Pi}
		\begin{bmatrix} \delta z_k\\ \delta v_k \end{bmatrix}
		\label{eq:augmented_dissipation}
	\end{equation}
	for all $k\geq 0$ where $\delta z_k = \delta \hat{x}_k$ (the neural network has input $\hat{x}_k$) and $Q = -R_\phi$, $S = -S_\phi^\top$, $R = -Q_\phi$.
\end{proposition}

\begin{proof}
	Since $\phi$ is memoryless, the pair $(\delta z_k, \delta v_k)$ is determined by $z_k$ and $z'_k$ alone, so the set of admissible pairs is the same for every $k$.
	Definition~\ref{def:iqc} with $T = 0$ therefore gives the pointwise inequality
	\begin{equation}
		\begin{bmatrix} \delta v_k \\ \delta z_k \end{bmatrix}^\top \Pi_\phi \begin{bmatrix} \delta v_k \\ \delta z_k \end{bmatrix} \geq 0, \quad \forall\, k\geq 0.
		\label{eq:iqc_pointwise}
	\end{equation}
	With this $\Pi$, the supply term of~\eqref{eq:augmented_dissipation} is then nonpositive at every $k$, so that $V(\delta \hat{x}_{k+1})\leq\beta\,V(\delta \hat{x}_k)$ and, iterating over $k$, $V(\delta \hat{x}_k)\leq\beta^k V(\delta \hat{x}_0)$.
	Since $\lambda_{\min}(P)\lVert\delta \hat{x}_k\rVert^2 \leq V(\delta \hat{x}_k)$ and $V(\delta \hat{x}_0) \leq \lambda_{\max}(P)\lVert\delta \hat{x}_0\rVert^2$, this yields
	\begin{equation}
		\lVert\delta \hat{x}_k\rVert^2 \leq \cond(P)\;\beta^{k}\,\lVert\delta \hat{x}_0\rVert^2,
	\end{equation}
	so the observer is contracting at rate $\beta$.
\end{proof}
\vspace{1mm}
\begin{remark}
	\label{rem:beta_rho_tradeoff}
	In equation~\eqref{eq:augmented_dissipation}, the multiplier term tightens the dissipation inequality, since the storage must decrease at rate $\beta$ while simultaneously absorbing the quadratic contribution given by $\Pi$.
	For a fixed $P$, demanding a faster contraction rate (smaller $\beta$) leaves less margin to accommodate the multiplier, which restricts the admissible operator class $\bm{\Phi}(\Pi_\phi)$.
	Conversely, enlarging $\bm{\Phi}(\Pi_\phi)$ to admit more expressive operators requires accepting a slower contraction rate.
	This tension is the $(\beta,\rho)$ trade-off formalized in Section~\ref{sec:pareto}.
\end{remark}

\subsection{Contraction via Dissipation LMIs}

The following result allows the efficient verification of condition~\eqref{eq:augmented_dissipation} \cite{seiler_stability_2015,hu_robustness_2017,manchester_robust_2018}.

\begin{theorem}
	\label{thm:diss_lmi}
	Condition~\eqref{eq:augmented_dissipation} holds if there exist $P\in\Sbb^n_+$ and a scalar $\beta\in(0,1]$ such that
	\begin{equation}
		\begin{bmatrix}
			F(\xi)^\top P\, F(\xi) - \beta P - Q & F(\xi)^\top P - S \\
			P\,F(\xi) - S^\top & P - R
		\end{bmatrix}
		\prec 0
		\label{eq:main_lmi}
	\end{equation}
	holds for all $\xi\in\R^n$, where $\xi$ is a free parameter not tied to any particular trajectory.
\end{theorem}

\begin{proof}
	Take the differential dynamics~\eqref{eq:variational} with $\delta z_k = \delta\hat{x}_k$, and leave $\delta v_k$ free, since the multiplier constrains it for every operator in $\bm{\Phi}(\Pi_\phi)$ (Definition~\ref{def:iqc} holds for all increments, hence for the differentials in the limit $\delta z_k \to 0$) rather than just for one realization of $\phi_\theta$.
	Substituting into~\eqref{eq:augmented_dissipation} and collecting all terms gives
	\begin{equation}
	\begin{split}
		&V(\delta\hat{x}_{k+1}) - \beta\,V(\delta\hat{x}_k) - \begin{bmatrix}\delta z_k\\ \delta v_k\end{bmatrix}^\top \Pi \begin{bmatrix}\delta z_k\\ \delta v_k\end{bmatrix}\\
		&\qquad = \begin{bmatrix}\delta\hat{x}_k\\ \delta v_k\end{bmatrix}^\top \mathcal{M}\bigl(\hat{x}_k\bigr) \begin{bmatrix}\delta\hat{x}_k\\ \delta v_k\end{bmatrix},
	\end{split}
	\end{equation}
	where $\mathcal{M}(\xi)$ denotes the matrix on the left-hand side of~\eqref{eq:main_lmi}.
	Since~\eqref{eq:main_lmi} holds for every $\xi\in\R^n$, it holds also at $\xi = \hat{x}_k$, and~\eqref{eq:augmented_dissipation} follows.
\end{proof}

Because $\xi$ ranges over all of $\R^n$, the resulting contraction guarantee is global, and holds for every pair of observer trajectories.
The guarantee can be restricted to a local one by restricting $\xi$ to a smaller set.

\begin{remark}
	When $f$ and $h$ are polynomial, $F(\xi)$ is a polynomial matrix in $\xi$, and the pointwise dissipation condition~\eqref{eq:main_lmi} can be verified via sum-of-squares (SOS) programming~\cite{manchester_control_2017} at the cost of an SDP whose size grows also with the degree of $f$ and $h$.
	In the linear case $f(x) = A_{\mathrm{ol}}\,x$, $h(x) = C_y\,x$, the Jacobian reduces to the constant matrix $F = A_{\mathrm{ol}} - LC_y$ and the pointwise condition collapses to a single LMI.
	For simplicity, the experiments of Section~\ref{sec:examples} are confined to the linear case.
\end{remark}

Even when $F = A_{\mathrm{ol}} - LC_y$ is constant, the condition~\eqref{eq:main_lmi} is bilinear in $(P,L)$ through the product $F^\top P\,F$.
Via a standard Schur complement and the change of variables $Y = PL$, it can be equivalently written as
\begin{equation}
	\begin{bmatrix}
		\beta P + Q & S & (PA_{\mathrm{ol}} - YC_y)^\top \\
		S^\top & R & P \\
		PA_{\mathrm{ol}} - YC_y & P & P
	\end{bmatrix}
	\succ 0,
	\label{eq:schur_lmi}
\end{equation}
which is \emph{affine} in $(P,Y,Q,S,R)$ for fixed $\beta$.
We call a tuple $(P,Y,Q,S,R)$, with $P\succ 0, Q\prec 0, R \succ 0$, that solves LMI~\eqref{eq:schur_lmi} a \emph{certificate}.
Note that, for every certificate that contains $(Q,S,R)$, we can obtain the corresponding multiplier $\Pi_\phi$ by setting $Q_\phi = -R, S_\phi = -S^\top, R_\phi = -Q$.
This multiplier can then be used in the REN implementation of $\phi_\theta$ to ensure that the observer~\eqref{eq:observer} is contractive for any $\theta\in\mathbb{R}^{n_\theta}$.
The observer gain is recovered as $L = P^{-1}Y$.

\begin{remark}
	\label{rem:other_parameters}
	The step that turns $L$ from a design input into a decision variable is the change of variables $Y = PL$.
	In~\eqref{eq:schur_lmi} the gain appears only through its product with $P$, so that product can be renamed as a single decision variable and $L$ recovered afterwards as $P^{-1}Y$.
	This is a standard linearizing method of LMI synthesis~\cite{scherer_multiobjective_1997}, and it applies to any design parameter that a congruence can reduce to a single product with the Lyapunov matrix or its inverse.
\end{remark}

\subsection{The \texorpdfstring{$(\beta,\rho)$}{(beta, rho)} Pareto Front}
\label{sec:pareto}

For a given plant $(f, h)$, the pointwise condition~\eqref{eq:main_lmi} (or its Schur complement form~\eqref{eq:schur_lmi}) defines a feasibility region in the space of certificates $(P,Y,Q,S,R)$ parametrized by $\beta$.
The term in condition~\eqref{eq:augmented_dissipation} due to the neural operator $\phi_\theta$ is a quadratic function of $\delta \hat{x}_k$ and $\delta v_k$.
We capture the ``size'' of this term by the maximum eigenvalue of matrix $Q$, which bounds the energy $\phi_\theta$ may inject in the state estimate $\hat{x}_k$.
The admissible gain $\rho$, for fixed $\beta$, is then computed as the solution of the following SDP
\begin{equation}
	\begin{array}{rl}
		\underset{P,Y,Q,S,R,\rho^2}{\text{maximize}} & \rho^2 \\
		\text{subject to} & \text{LMI}~\eqref{eq:schur_lmi},\ P\succ 0 \\
		& Q \preccurlyeq -\rho^2 I,\ R \succ 0,\ \tr(R) = n,
	\end{array}
	\label{eq:pareto_sdp}
\end{equation}
where the normalization $\tr(R) = n$ is necessary because the IQC condition~\eqref{eq:iqc_condition} is homogeneous in $\Pi_\phi$, equivalently in $\Pi$.
Without it, $\lambda_{\max}(Q)$ can be made arbitrarily negative by joint positive rescaling of $(Q,S,R)$ and $P$, without enlarging the operator class $\bm{\Phi}(\Pi_\phi)$.
Fixing $\tr(R)=n$ fixes the scale of $\Pi$ so that it represents a non-trivial multiplier and $\rho^2$ reflects the size of the class $\bm{\Phi}(\Pi_\phi)$.

\begin{figure}[t]
	\centering
	\includegraphics[width=\columnwidth]{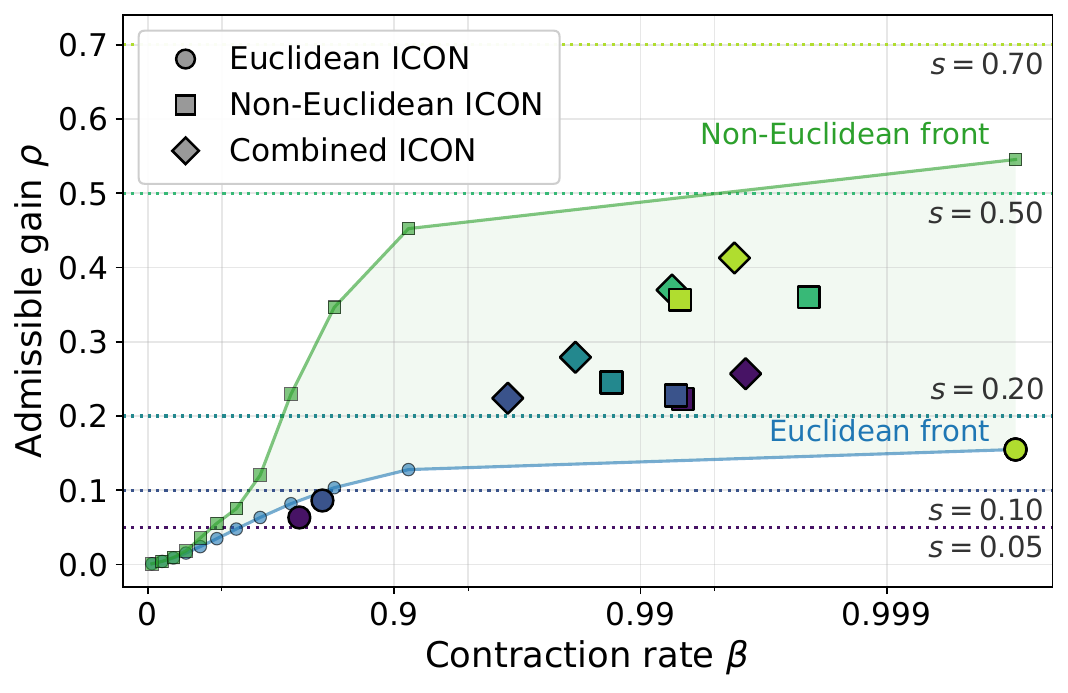}
	\caption{Pareto fronts of the plant~\eqref{eq:system_example} against the contraction rate $\beta$, on an axis linear in $-\log(1-\beta)$, with the learned operating points overlaid. The small circles are the $12$ sampled Euclidean (Lipschitz) certificates and the small squares are the non-Euclidean ones. The heavy markers are the medians of the learned operating points over the $5$ initializations, one per target gain and class, circles for Euclidean, squares for non-Euclidean, diamonds for combined, each colored with the target gain $s$ it was trained for. The Euclidean medians for $s \geq 0.2$ coincide, as do the non-Euclidean ones for $s \leq 0.1$.}
	\label{fig:pareto}
\end{figure}

Since the SDP jointly optimizes over the observer gain (through $Y$) and the IQC multiplier, the resulting Pareto front reflects the best achievable trade-off when $L$ is co-designed with $(Q,S,R)$.
Varying $\beta$ over $[\beta_{\min},1]$, where $\beta_{\min}$ is the smallest feasible contraction rate, traces the \emph{Pareto front} $\mathcal{P} = \{(\beta, \rho^*(\beta))\}$, as illustrated in Fig.~\ref{fig:pareto}.

\begin{remark}
	When the certificate is restricted to $Q = -\rho^2 I$, $S=0$, $R=I$ (Euclidean or Lipschitz certificate), the resulting front is typically significantly more conservative than the general $(Q,S,R)$ front.
	The gap is particularly pronounced for systems with anisotropic dynamics, i.e., systems whose eigenvalues have different magnitudes or whose eigenvectors are misaligned with the coordinate axes.
\end{remark}

\section{Differentiable IQC Multiplier Parametrization}
\label{sec:certificate_interp}

The main challenge in learning IQC multipliers and observer gains is to parametrize them in a way that is both (i)~differentiable, to enable gradient-based optimization, and (ii)~feasibility-preserving, so that the dissipation LMI is satisfied for \emph{all} parameter values, implying contraction of the observer, without requiring online verification.
We now describe a construction that achieves both properties simultaneously.

\subsection{Sampled Certificate Set}

Since each point $(\beta,\rho)$ on and below the Pareto front corresponds to a different feasible certificate, sampling a finite grid of points provides a discrete set of LMI solutions that spans the available trade-off between convergence speed and admissible gain.
This sampling is a one-time offline cost, after which the certificates are available for interpolation during training of $\phi_\theta$.
The (locally) optimal interpolation is learned from data, alongside the parameters $\theta$ of the network.
The grid density $N$ controls the resolution at which the optimizer of $\phi$ can navigate the area enclosed by the $(\beta,\rho)$ front.
A coarser grid restricts the learnable certificates to a smaller polytope, while a finer grid enlarges it at the expense of additional SDP solves and slower convergence during training.

Let $\beta_1 < \ldots < \beta_N$ be a finite grid of contraction rates in $[\beta_{\min}, \beta_{\max}]$ with $\beta_N = \beta_{\max}$, and for each $\beta_i$, let $\mathcal{C}_i = (P_i, Y_i, Q_i, S_i, R_i)$ be a solution to the SDP~\eqref{eq:pareto_sdp}.
The collection $\mathfrak{C} = \{\mathcal{C}_i\}_{i=1}^N$ constitutes the \emph{sampled certificate set}.
The two endpoints have different origins.
The lower one, $\beta_{\min}$, is the fastest rate at which~\eqref{eq:pareto_sdp} remains feasible, and is a property of the plant and of any constraint added to~\eqref{eq:pareto_sdp}.
The upper one, $\beta_{\max}$, is the slowest contraction rate the designer is willing to accept, and is therefore a specification.

By construction, each $\mathcal{C}_i$ satisfies the Schur complement LMI~\eqref{eq:schur_lmi} with parameters $(\beta_i, \rho_i)$.
Let $\mathcal{S}(\beta,P,Y,Q,S,R)$ denote the left-hand side of~\eqref{eq:schur_lmi}.
For fixed $\beta$ it is affine in $(P,Y,Q,S,R)$, which is the property that allows convex interpolation of its solutions.

\begin{lemma}
	\label{lem:convexity}
	Let $\mathcal{C}_i = (P_i, Y_i, Q_i, S_i, R_i)$, $i\in\{1,\ldots,N\}$, be certificates satisfying the LMI~\eqref{eq:schur_lmi} for respective contraction rates $\beta_i$.
	Let $\alpha\in\R^N$ be a vector in the probability simplex, i.e., $\alpha_i\geq 0$ and $\sum_i \alpha_i = 1$.
	Define
	\begin{equation}
        \begin{split}
            \bar{P} &= \textstyle\sum_{i} \alpha_i P_i, \quad \bar{Y} = \textstyle\sum_{i} \alpha_i Y_i,\\
            \bar{Q} &= \textstyle\sum_{i} \alpha_i Q_i, \quad \bar{S} = \textstyle\sum_{i} \alpha_i S_i, \quad \bar{R} = \textstyle\sum_{i} \alpha_i R_i.
		      \label{eq:convex_comb}
        \end{split}
	\end{equation}
	Then $\bar{\mathcal{C}} \coloneqq (\bar{P}, \bar{Y}, \bar{Q}, \bar{S}, \bar{R})$ satisfies the LMI~\eqref{eq:schur_lmi} with a contraction rate $\beta^* \leq \beta_{\max} = \max_i \beta_i$, and the interpolated observer gain $\bar{L} = \bar{P}^{-1}\bar{Y}$ is well-defined.
\end{lemma}

\begin{proof}
	Since each $P_i \succ 0$ and $\beta_{\max} \geq \beta_i$, we have $(\beta_{\max} - \beta_i)\,\diag(P_i,0,0) \succcurlyeq 0$, so each certificate also satisfies the LMI at rate $\beta_{\max}$: $\mathcal{S}(\beta_{\max}, \mathcal{C}_i) \succcurlyeq \mathcal{S}(\beta_i, \mathcal{C}_i) \succ 0$.
	For fixed $\beta_{\max}$, the map $\mathcal{C} \mapsto \mathcal{S}(\beta_{\max}, \mathcal{C})$ is affine, so convexity of the positive definite cone gives $\mathcal{S}(\beta_{\max}, \bar{\mathcal{C}}) = \sum_i \alpha_i\, \mathcal{S}(\beta_{\max}, \mathcal{C}_i) \succ 0$.
	Since the LMI is monotone in $\beta$, the smallest feasible rate $\beta^*$ satisfies $\beta^* \leq \beta_{\max}$.
	Finally, each $P_i \succ 0$ and $\alpha_i \geq 0$ with $\sum_i \alpha_i = 1$ imply $\bar{P} \succ 0$, so $\bar{L} = \bar{P}^{-1}\bar{Y}$ is well-defined.
\end{proof}

\begin{remark}
	\label{rem:beta_conservatism}
	The rate $\beta_{\max}$ is a sufficient, but generally conservative, contraction rate for the interpolated certificate.
    Strict inequality ($\beta^* < \beta_{\max}$) may hold when $\alpha_i > 0$ for some $i$ with $\beta_i < \beta_{\max}$, as those terms contribute additional positive definiteness slack.
    After training, $\bar{\mathcal{C}}$ is fixed and the LMI is linear in the single scalar $\beta$, so the smallest feasible rate $\beta^*$ can be efficiently recovered by bisection or, equivalently, via a generalized eigenvalue problem.
    Since $\beta^* \leq \beta_{\max}$ holds for every $\alpha$ in the simplex, the specification encoded in the sampling acts as a limit that cannot be crossed, and what is admissible for a given application is fixed offline while learning only places the operating point inside it.
\end{remark}

\subsection{Smooth Simplex Mapping}

The probability simplex $\Delta_N = \{\alpha\in\R^N : \alpha_i\geq 0,\,\sum_i\alpha_i = 1\}$ is a constrained set.
To enable unconstrained optimization over the interpolation weights, we employ the standard smooth surjection
\begin{equation}
	\alpha_i(\zeta) = \frac{e^{\zeta_i}}{\sum_{j=1}^N e^{\zeta_j}}, \quad \zeta\in\R^N,
	\label{eq:softmax}
\end{equation}
which maps $\R^N$ onto the interior of $\Delta_N$.
The mapping $\zeta\mapsto\alpha(\zeta)$ is infinitely differentiable, and its Jacobian has rank $N-1$ everywhere.
Crucially, $\alpha(\zeta) > 0$ componentwise and $\lVert \alpha(\zeta) \rVert_1 = 1$ for all $\zeta\in\R^N$, so the interpolated certificate is always a strictly positive convex combination.

\section{Main Result}
\label{sec:main_result}

We now state the main result, which establishes that both the IQC multiplier and the observer gain can be learned via unconstrained optimization while maintaining certified LMI feasibility throughout training.

\begin{proposition}
	\label{prop:main}
	Let $\mathfrak{C} = \{\mathcal{C}_i\}_{i=1}^N$ be a sampled certificate set, where each $\mathcal{C}_i = (P_i,Y_i,Q_i,S_i,R_i)$ satisfies the Schur complement LMI~\eqref{eq:schur_lmi} with contraction rate $\beta_i \in [\beta_{\min},\beta_{\max}]$ and admissible gain $\rho_i > 0$.
	Define the parametric certificate
	\begin{equation}
		\mathcal{C}(\zeta) = \bigl(\bar{P}(\zeta),\, \bar{Y}(\zeta),\, \bar{Q}(\zeta),\, \bar{S}(\zeta),\, \bar{R}(\zeta)\bigr), \quad \zeta\in\mathbb{R}^N,
		\label{eq:parametric_cert}
	\end{equation}
	where each component is given by~\eqref{eq:convex_comb} with $\alpha(\zeta)$ defined by~\eqref{eq:softmax}.

	Then, for every $\zeta\in\R^N$, the certificate $\mathcal{C}(\zeta)$ satisfies:
	\begin{enumerate}
		\item[(i)] \textbf{LMI feasibility:}\\
        $\mathcal{S}\bigl(\beta^*(\zeta),\, \bar{P}(\zeta),\, \bar{Y}(\zeta),\, \bar{Q}(\zeta),\, \bar{S}(\zeta),\, \bar{R}(\zeta)\bigr) \succ 0$ for some $\beta^*(\zeta) \leq \beta_{\max} = \max_i \beta_i$.

		\item[(ii)] \textbf{Positive definiteness:} $\bar{P}(\zeta) \succ 0$.

		\item[(iii)] \textbf{Multiplier inertia:} $\bar{Q}(\zeta) \prec 0$ and $\bar{R}(\zeta) \succ 0$.

		\item[(iv)] \textbf{Smoothness:}\\
        The maps $\zeta \mapsto \mathcal{C}(\zeta)$ and $\zeta \mapsto \bar{L}(\zeta) = \bar{P}(\zeta)^{-1}\bar{Y}(\zeta)$ are infinitely differentiable.
	\end{enumerate}
\end{proposition}

\begin{proof}
	\emph{(i)} follows directly from Lemma~\ref{lem:convexity} and the fact that $\alpha(\zeta)$ lies in the probability simplex for all $\zeta$.

	\emph{(ii)} Since each $P_i\succ 0$ and $\alpha_i(\zeta)>0$ for all $\zeta$, we have $\bar{P}(\zeta) = \sum_i \alpha_i(\zeta)P_i \succ 0$ as a strictly positive combination of positive definite matrices.

	\emph{(iii)} Similarly, each $Q_i \prec 0$ implies $\bar{Q}(\zeta) \prec 0$, and each $R_i\succ 0$ implies $\bar{R}(\zeta)\succ 0$.

	\emph{(iv)} The map $\zeta\mapsto\alpha(\zeta)$ is $C^\infty$, and the certificate components are linear in $\alpha$, hence $C^\infty$ in $\zeta$.
	The map $P\mapsto P^{-1}$ is $C^\infty$ on the positive definite cone, so $\bar{L}(\zeta) = \bar{P}(\zeta)^{-1}\bar{Y}(\zeta)$ is $C^\infty$ as a composition of smooth maps.
\end{proof}

\begin{remark}
	Proposition~\ref{prop:main} transforms the IQC learning problem into an unconstrained optimization over $\zeta\in\R^N$.
	The map $\zeta\mapsto\mathcal{C}(\zeta)$ is smooth and feasibility-preserving, so the multiplier and gain can be updated via gradient descent jointly with the parameters $\theta$ of $\phi$, without any online SDP solve.
	This is different from the fixed-multiplier setting of~\cite{revay_recurrent_2024,wang_youla-ren_2022}, where the IQC is a design input rather than a learned quantity.
\end{remark}

\section{The ICON Method}
\label{sec:icon_method}

We now assemble the learned IQC multiplier $\Pi_\phi$ and observer gain from Section~\ref{sec:main_result} with the REN operator $\phi_\theta$ to produce a jointly trainable system with certified contraction, which we call ICON (Incrementally Constrained Observer with Neural dynamics).

\subsection{Pareto Front Sampling}
\label{sec:front_sampling}

Given the plant $(f,h)$, the certificate set $\mathfrak{C} = \{\mathcal{C}_i\}_{i=1}^N$ with $\mathcal{C}_i = (P_i,Y_i,Q_i,S_i,R_i)$ is obtained by solving the SDP~\eqref{eq:pareto_sdp} for a grid of contraction rates $\beta_1 < \ldots < \beta_N$ using a standard interior-point solver (e.g., MOSEK~\cite{mosek}).
The observer gain for each certificate is $L_i = P_i^{-1}Y_i$.
Two classes of IQC certificates can be sampled: \emph{Euclidean} ($Q_i = -\rho_i^2 I$, $S_i = 0$, $R_i = I$), corresponding to Lipschitz bounds, and \emph{non-Euclidean} with general ($Q_i \prec 0, S_i, R_i \succ 0$), which capture a more expressive directional structure at a higher computational cost.
Both types can be combined in a single set $\mathfrak{C}$.
The multiplier is defined only up to a positive scaling, so the non-Euclidean class requires a normalization to render $\rho_i$ bounded, for which we impose $\tr(R_i) = n$.
This is also satisfied by the Euclidean choice $R_i = I$.
For Euclidean certificates $\rho_i$ is exactly the admissible Lipschitz constant, whereas for non-Euclidean certificates it is the gain the multiplier admits in every direction.

\subsection{Co-Learning of Operator and Multiplier}

By design, it is possible to enforce an incremental IQC on a REN operator via the multiplier $\Pi_\phi$.
REN operators are also differentiable with respect to the parameters $\theta$, and, crucially, with respect to the enforced multiplier $\Pi_\phi$ \cite{revay_recurrent_2024}.
We connect this to our parametrization of Proposition~\ref{prop:main} by enforcing on the operator $\phi_\theta$ the multiplier $\Pi_\phi(\zeta)$ composed of $Q_\phi = -\bar{R}(\zeta),\, S_\phi = -\bar{S}(\zeta)^\top,\, R_\phi = -\bar{Q}(\zeta)$.
This leads to an extended parametrization of the operator, which we indicate with $\phi_{(\theta,\zeta)}$.
The learnable parameter vector is then $(\theta, \zeta)\in\R^{n_\theta + N}$, where $\theta$ controls \emph{what the nonlinear operator does}, and $\zeta$ simultaneously controls \emph{what IQC it satisfies} and \emph{which observer gain is used}.

\begin{proposition}
	\label{prop:icon_diss}
	For every $(\theta, \zeta)\in\R^{n_\theta+N}$, $\phi_{(\theta,\zeta)}$ is differentiable with respect to $(\theta,\zeta)$ and the observer~\eqref{eq:observer} with neural operator $\phi_{(\theta,\zeta)}$ and gain $\bar{L}(\zeta)$ is contracting at rate $\beta^*(\zeta) \leq \beta_{\max}$.
\end{proposition}

\begin{proof}
	REN operators are differentiable with respect to both $\theta$, the unconstrained parameters, and $\Pi_\phi$, the enforced IQC multiplier.
	Differentiability with respect to $\zeta$ then follows from the assignment $Q_\phi = -\bar{R}(\zeta),\, S_\phi = -\bar{S}(\zeta)^\top,\, R_\phi = -\bar{Q}(\zeta)$, and differentiability of $(\bar{Q}(\zeta),\bar{S}(\zeta),\bar{R}(\zeta))$ with respect to $\zeta$, proved in Proposition~\ref{prop:main}.
	The same proposition also ensures that~\eqref{eq:schur_lmi} holds at $\beta^*(\zeta)$ for any $\zeta$, which is equivalent to~\eqref{eq:main_lmi}, so Theorem~\ref{thm:diss_lmi} gives~\eqref{eq:augmented_dissipation} and Proposition~\ref{prop:full_contraction} the contraction of the observer.
\end{proof}

The optimization problem with cost function $J$,
\begin{equation}
	\min_{\theta\in\R^{n_\theta},\,\zeta\in\R^N} J(\theta, \zeta)
	\label{eq:training}
\end{equation}
is \emph{unconstrained} and can thus be solved (to local optimality) with a large number of standard gradient-based optimizers (e.g., Adam~\cite{kingma_adam_2014}).

The only structural requirement on $J$ is differentiability in $(\theta,\zeta)$, since Proposition~\ref{prop:icon_diss} holds at every parameter value and Lemma~\ref{lem:convexity} puts the upper limit on the certified contraction rate $\beta_{\max}$ regardless of the objective.

The complete ICON method is summarized in Algorithm~\ref{alg:icon}.

\begin{algorithm}[t]
	\caption{ICON: Incrementally Constrained Observer with Neural dynamics}
	\label{alg:icon}
	\begin{algorithmic}[1]
		\REQUIRE Plant $(f, h)$, grid $\{\beta_i\}_{i=1}^N$, training data $\mathcal{D}$ containing input-output samples of $g$.
		\STATE \textbf{Offline:} For each $\beta_i$, solve SDP~\eqref{eq:pareto_sdp} $\to$ $\mathcal{C}_i = (P_i,Y_i,Q_i,S_i,R_i)$
		\STATE Initialize $\theta \sim \mathcal{N}(0,\sigma_\theta^2 I)$, $\zeta \gets \log\alpha$ with $\alpha \sim \mathrm{Unif}(\Delta_N)$
		\FOR{epoch $= 1,2,\ldots$ and mini-batch $\mathcal{B} \subset \mathcal{D}$}
		\STATE Compute $\alpha(\zeta)$ via~\eqref{eq:softmax}
		\STATE Interpolate\\
        $(\bar{P},\bar{Y},\bar{Q},\bar{S},\bar{R}) \gets \sum_i \alpha_i (P_i,Y_i,Q_i,S_i,R_i)$
		\STATE Set $\Pi_\phi$ on the REN operator
		\STATE Evaluate $J(\theta,\zeta)$ on $\mathcal{B}$
		\STATE Update $(\theta,\zeta) \gets \text{Adam}\bigl((\theta,\zeta),\, \nabla J\bigr)$
		\ENDFOR
		\ENSURE Trained model $\phi_{(\theta^*,\zeta^*)}$, learned gain $\bar{L}^*$
	\end{algorithmic}
\end{algorithm}

\section{Numerical Experiment}
\label{sec:examples}

We illustrate the framework on a linear plant, the simplest instance of~\eqref{eq:plant} with $f(x) = A_{\mathrm{ol}}\,x$.
We consider a $2$-dimensional system with anisotropic dynamics and partial observation:
\begin{equation}
	A_{\mathrm{ol}} = R_\vartheta \begin{bmatrix} 0.85 & 0.3 \\ 0 & 0.65 \end{bmatrix} R_\vartheta^\top, \quad C_y = [1,\;0],
	\label{eq:system_example}
\end{equation}
where $R_\vartheta$ is a rotation matrix with $\vartheta = 60^\circ$.
This system is Schur stable (eigenvalues $0.85$ and $0.65$), but strongly non-normal, since the off-diagonal shear and the $60^\circ$ rotation misalign the eigenvectors from the observation axis, creating directional structure that is poorly captured by Lipschitz bounds.
The unknown part of the dynamics is a rotated saturating nonlinearity with a linear offset, $g(x) = A_g x + \tfrac{1}{60} R_\vartheta\,[2\tanh(3\,r_1),\; 0.1\tanh(r_2)]^\top$, $A_g = R_\vartheta\,\mathrm{diag}(-s,\,-s/7)\,R_\vartheta^\top$, $[r_1,\,r_2]^\top = R_\vartheta^\top x$, whose slope along $r_1$ spans $[-s,\,-s+0.1]$, so the target gain is controlled by $s$.

The observer gain $L$ is co-designed with the IQC certificate via the Schur complement LMI~\eqref{eq:schur_lmi}, with the gain constraint $\gamma^2 P \succcurlyeq Y Y^\top$ added to the SDP~\eqref{eq:pareto_sdp}, which limits the transmission gain of measurement noise into the estimation error and bounds the gain magnitude by $\lVert L\rVert \leq \gamma/\sqrt{\lambda_{\min}(P)}$.
In practice, such a limit would be dictated by the requirements of the application rather than obtained by tuning.
We fix $\gamma = 0.08$ and, unless stated otherwise, $s = 0.7$ for every model in this section.
Training data consists of $90$ trajectories of $250$ steps of the plant~\eqref{eq:plant} with two terms added to the state equation, a known excitation $u_k$ that is also fed to the observer and process noise $w_k \sim \mathcal{N}(0,\,0.25\,I)$, divided into blocks of $50$ steps and split by trajectory $310/70/70$ into train/validation/test sets.

\subsection{Pareto Front with Co-Designed Observer Gain}

Fig.~\ref{fig:pareto} shows the Pareto front obtained by solving the SDP~\eqref{eq:pareto_sdp} for both Euclidean and non-Euclidean certificates.
Each point on the front corresponds to a jointly optimized certificate $(P, Y, Q, S, R)$ with observer gain $L = P^{-1}Y$.
Along the front, Euclidean certificates (circles) need larger gains to compensate for their Lipschitz constraint, $\|L\| = 0.252$ for $\rho = 0.155$ at $\beta = 1$, while non-Euclidean ones (squares) reach higher $\rho$ with smaller gains, $\|L\| = 0.117$ for $\rho = 0.545$ at the same $\beta$.
At $\beta = 1$, the non-Euclidean front reaches an admissible gain $3.5\times$ larger than the Euclidean one, and the advantage grows with $\beta$.
The two fronts coincide at the smallest feasible rate and separate monotonically thereafter, since a faster contraction rate leaves little room for directional structure to be exploited.
On the other hand, searching over Euclidean (or Lipschitz) certificates leads to fewer decision variables in the SDP~\eqref{eq:pareto_sdp}, making it faster to compute during the certificate sampling stage.
The sampled rates for the experiment are $12$ values equally spaced in $\beta \in [0.039,\, 1]$, so that $\beta_{\max} = 1$, the widest possible interval, which spans the entire front rather than reflecting a design requirement.

\subsection{Comparative Training}

We train five models: a \emph{Euclidean ICON} interpolating $12$ Euclidean certificates ($Q = -\rho^2 I$, $S=0$, $R=I$) and a \emph{non-Euclidean ICON} interpolating $12$ general $(Q,S,R)$ certificates, both with $172$ parameters, a \emph{combined ICON} over all $24$ certificates with $184$ parameters, an \emph{unconstrained MLP} of two layers of $10$ $\tanh$ memoryless units with $162$ parameters and no dissipation constraint, and a \emph{wide unconstrained MLP} with $50$ units per layer and $2802$ parameters.
All ICON models use $n_w = 10$ nonlinear channels, and every model is trained for $6000$ epochs with Adam on batches of $1024$ samples, minimizing the MSE between the predicted and true $g$, and retaining the parameters that achieve the lowest validation loss.
The learning rate was set to $0.03$ for ICON and to $10^{-3}$ for the MLPs, the latter selected by a sweep over seven rates on the MLP baseline.
Each configuration is trained from $5$ random initializations of $(\theta, \zeta)$, with the data and the certificate sets held fixed, and all reported quantities are means $\pm$ standard deviations over those runs.

The MSE is one instance of $J$ among many that drive $\phi$ towards $g$ by penalizing the residual $d_k = g(x_k) - \phi(x_k)$.
Other potential choices are a max-norm residual and a residual weighted by the interpolated metric $\bar{P}(\zeta)$.
Terms acting on $\zeta$ alone, such as a penalty on $\lVert\bar{L}(\zeta)\rVert$ that limits noise sensitivity, can be added to any of these.

\begin{figure}[t]
	\centering
	\includegraphics[width=\columnwidth]{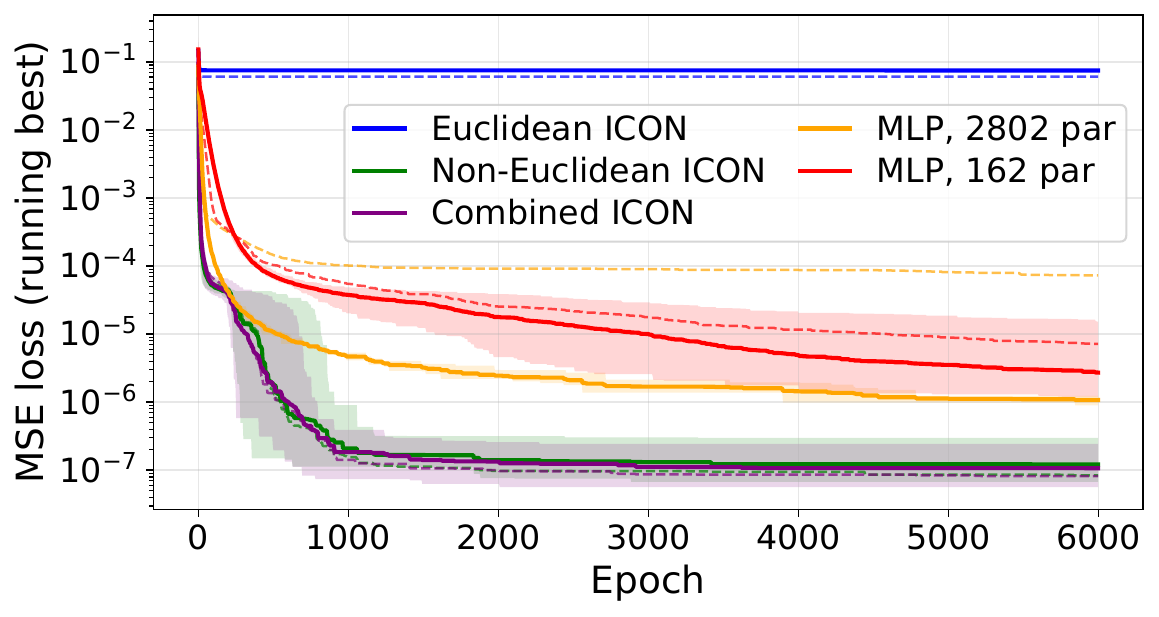}
	\caption{Training and validation loss at $s = 0.7$. The lines are geometric means over $5$ initializations of the running minimum over epochs. Solid lines are training losses, dashed lines validation losses, shaded regions the training-loss envelopes over the $5$ initializations. The Euclidean ICON (blue) plateaus, while the non-Euclidean (green) and combined (purple) variants reach losses five orders of magnitude lower. The wide (orange) and matched (red) unconstrained MLPs stop two to three orders of magnitude above the certified non-Euclidean models in validation loss.}
	\label{fig:learning}
\end{figure}

Fig.~\ref{fig:learning} shows the training and validation losses.
The non-Euclidean and combined ICON models converge to test losses of $(1.06 \pm 0.35)\cdot 10^{-7}$ and $(0.92 \pm 0.34)\cdot 10^{-7}$, respectively, while the Euclidean ICON plateaus at $(7.49 \pm 0.00)\cdot 10^{-2}$, more than five orders of magnitude worse.
The separation persists across initializations, since even the best Euclidean run (test loss $7.48\cdot 10^{-2}$) remains far above the worst run of either other variant (test loss $1.5\cdot 10^{-7}$).
This large gap arises because the Euclidean constraint ($Q = -\rho^2 I$, $R = I$) caps the incremental gain of $\phi$ at $\rho = 0.155$ in every direction, below the slope $s = 0.7$ of $g$ along $r_1$, so the Euclidean model cannot represent $g$ at all.
The non-Euclidean and combined models, on the other hand, differ only slightly, as their means are separated by less than one standard deviation.
Enlarging the certificate set therefore brings little benefit in fit, although the combined model reaches the lowest estimation error in Section~\ref{sec:observer_perf} and, containing both families, can by construction do no worse than either taken individually.
The two unconstrained baselines attain higher test losses, $(1.5 \pm 2.0)\cdot 10^{-5}$ for the matched MLP and $(5.9 \pm 1.6)\cdot 10^{-5}$ for the wide one, so here the certificate acts as a useful prior rather than a restrictive constraint.
That is not always the case as also shown by the ordering reversal in the tests that used the two lowest values of target gain, $s = 0.05$ and $s = 0.1$.

\subsection{Learned Certificates and Gain}

The advantage of the IQC learning formulation is that during training, by simple gradient descent, a $(\beta,\rho)$ trade-off is automatically selected from the task loss $J$ alone.
After training, each model's interpolation weights $\alpha(\zeta)$ determine a certificate $(P,Y,Q,S,R)$ from which the achieved trade-off is recovered.
The two coordinates are recovered separately: $\rho$ is determined by $\rho^2 = -\lambda_{\max}(\bar{Q}(\zeta))$; whereas $\beta^*$, the smallest contraction rate at which~\eqref{eq:main_lmi} remains negative definite, is recovered by bisection.
Since the bisection stops at the crossing into infeasibility, $\mathcal{M}$ has zero maximum eigenvalue at $\beta^*$ for every model up to the numerical tolerance, so the certificate is active there by construction.
Table~\ref{tab:learned} collects the outcome over the $5$ initializations.

The Euclidean model reaches in every initialization the certificate of its front with the highest $\rho = 0.155$, $\beta^* = 1$ with $\|L\| = 0.252$, while the non-Euclidean and combined models both settle at $\beta^* = 0.994$, with $\rho = 0.375$ and $\rho = 0.401$ and gains $\|L\| = 0.105$ and $\|L\| = 0.102$.
Fig.~\ref{fig:pareto} shows the learned certificate points for different gains of the unknown function $g$, namely $s \in \{0.05, 0.1, 0.2, 0.5, 0.7\}$.
The value of $\rho$ achieved by the Euclidean class rises with $s$ and saturates at $\rho = 0.155$ from $s \geq 0.2$, whereas the non-Euclidean and combined medians stay between $\rho = 0.22$ and $\rho = 0.28$ for $s \leq 0.2$ and rise only at $s = 0.5$, so at the lowest values of $s$ their certificate point is not set by the task.
This illustrates the main advantage of the non-Euclidean certificates: a Euclidean multiplier admits only the ball $\|\Delta\phi\| \leq \rho\,\|\Delta x\|$ of incremental gains, whereas the learned $(Q,S,R)$ admit the incremental gains that $g$ exhibits on the training domain, including the slope $s = 0.7$ along $r_1$, while their inscribed ball has radius only $\rho = 0.375$.

Fixing the multiplier to a single certificate, or, in other words, setting $N=1$ with no interpolation, leaves the outcome strongly dependent on which certificate is chosen, with test losses spanning more than five orders of magnitude across the six fixed certificates tested.
The multiplier with the highest $\rho$ comes within a factor of two of the interpolated model in test loss, $(1.9 \pm 0.6)\cdot 10^{-7}$, but only by exhausting the contraction budget at $\beta^* = 1$, whereas the interpolating model reaches its lower loss at $\beta^* = 0.994$.

\begin{table}[t]
	\caption{Learned certificates and observer performance at $s = 0.7$, mean $\pm$ standard deviation over $5$ random initializations, steady-state error in units of $10^{-3}$.}
	\label{tab:learned}
	\centering
	\footnotesize
	\renewcommand{\arraystretch}{1.1}
	\setlength{\tabcolsep}{3pt}
	\begin{tabular}{lccc}
		\hline
		& Euclidean & Non-Euclidean & Combined \\
		\hline
		Test MSE\rule{0pt}{2.6ex}      & $7.49\,10^{-2}$ & $(1.06{\pm}0.35)\,10^{-7}$ & $(0.92{\pm}0.34)\,10^{-7}$ \\
		$\beta^*$                  & $1.0$ & $0.994{\pm}0.002$ & $0.994{\pm}0.004$ \\
		$\rho$                         & $0.155$ & $0.375{\pm}0.030$ & $0.401{\pm}0.037$ \\
		$\|L\|$                        & $0.252$ & $0.105{\pm}0.007$ & $0.102{\pm}0.005$ \\
		Steady-state $\|e_k\|$         & $52.4{\pm}1.6$ & $1.9{\pm}0.2$ & $1.8{\pm}0.1$ \\
		\hline
	\end{tabular}
\end{table}

\addtolength{\textheight}{-1.4cm}
\subsection{Observer Performance under Measurement Noise}
\label{sec:observer_perf}

To assess the practical impact of the certificate choice, we deploy the trained observers with partial observation $C_y = [1,\;0]$ and additive Gaussian measurement noise ($\sigma = 0.021$).
Each observer uses its own learned gain $L = \bar{P}(\zeta)^{-1}\bar{Y}(\zeta)$.
The initial estimation error is set to $\|e_0\| = 14.14$.
Each of the $5$ trained models per certificate category is run over $20$ noise realizations, drawn once and shared across model evaluations so that every observer faces the same noise.
All observer quantities are then aggregated in two stages.
For a fixed model, the estimation error is first averaged over the $20$ realizations, and its steady-state error is taken to be the mean of $\|e_k\|$ over the last $100$ of $500$ steps.
Within a given certificate class, those per-model values are then further averaged across the $5$ random model initializations.
The reported standard deviations are due to variability across trained models rather than across noise realizations.

\begin{figure}[t]
	\centering
	\includegraphics[width=\columnwidth]{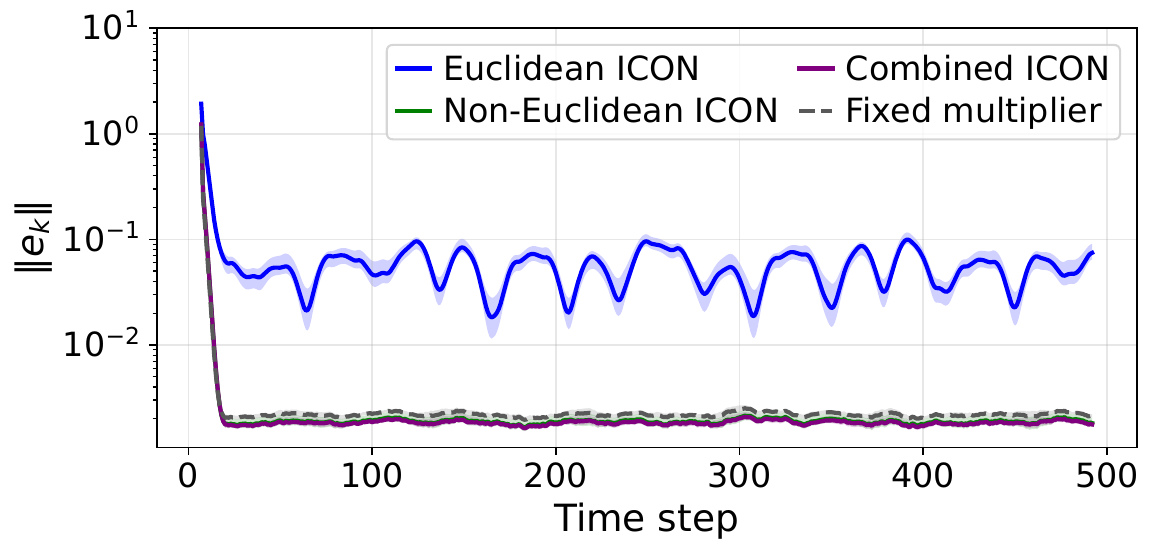}
    \caption{Aggregate observer estimation error at $s = 0.7$ under additive Gaussian measurement noise ($\sigma = 0.021$). Solid lines are windowed averages ($W\!=\!15$ steps) of the mean error, shaded bands are the envelopes across the $5$ models. The dashed gray trace is the fixed non-Euclidean multiplier with the highest admissible gain, $\rho = 0.545$.}
    \label{fig:observer}
\end{figure}

Fig.~\ref{fig:observer} compares the estimation performance.
The non-Euclidean and combined ICON observers achieve steady-state errors of $(1.9 \pm 0.2)\cdot 10^{-3}$ and $(1.8 \pm 0.1)\cdot 10^{-3}$, compared to $(52.4 \pm 1.6)\cdot 10^{-3}$ for the Euclidean variant.
As in the regression task, the separation holds across different initializations, since the worst non-Euclidean or combined run stays below the best Euclidean one.
The Euclidean model must compensate for its Lipschitz constraint with a larger gain, $\|L\| = 0.252$ against $\|L\| = 0.105$ and $\|L\| = 0.102$ for the other two variants.
This more aggressive gain amplifies the measurement noise, adding to the error caused by model mismatch, as the comparison in Fig.~\ref{fig:observer} shows.

\section{Conclusion}
\label{sec:conclusion}

We formulated and solved an IQC learning problem: learning which certificate a nonlinear operator should satisfy, jointly with the operator itself, so as to minimize a task objective while maintaining certified contraction.
The mechanism that enables this method is a differentiable parametrization of IQC multipliers via convex interpolation over precomputed certificates, which maintains LMI feasibility by construction and eliminates the need for online SDP solves.
It is also possible to co-learn the observer gain alongside the IQC multiplier in the observer design application via the LMI synthesis reformulation, which allows us to include the observer gain as part of the certificate.

The results demonstrate that learning the IQC and the observer gain can yield improvements over fixed Lipschitz certificates, both in approximation quality and observer estimation error.
This confirms that one source of conservatism in existing fixed-multiplier designs~\cite{revay_recurrent_2024,wang_youla-ren_2022} is the a priori choice of IQC, a choice that ICON delegates to the optimizer within a sampled certificate set.

Future work includes extensions to dynamic IQC multipliers, which would allow richer operator classes at the cost of additional filter states and larger LMI size, the treatment of operators with internal state, and the application to higher-dimensional problems.

\bibliographystyle{IEEEtran}
\bibliography{references}

\end{document}